\documentclass[useAMS,usenatbib]{mnras}
\usepackage{aas_macros,graphicx,times,multirow}

\usepackage{array}

\def \cxo{CXO\, J164710.2$-$455216}

\title[VLT observations of \cxo]{VLT observations of the magnetar \cxo\ and the detection of a candidate infrared counterpart\thanks{Based on observations collected at the European Organisation for 
Astronomical Research in the Southern Hemisphere under ESO programme 091.D-0071(B). }}
\author[V. Testa, R. P. Mignani, W. Hummel, et al.]
{\parbox{\textwidth}{V. Testa$^{1}$\thanks{E-mail: vincenzo.testa@oa-roma.inaf.it},
R. P. Mignani$^{2,3}$,
W. Hummel${^4}$,
N. Rea$^{5,6}$,
G. L. Israel$^{1}$
}
\\ \\
$^{1}$ INAF - Osservatorio Astronomico di Roma, via Frascati 33, 00078, Monte Porzio Catone, Italy \\
$^{2}$ INAF - Istituto di Astrofisica Spaziale e Fisica Cosmica Milano, via E. Bassini 15, 20133, Milano, Italy\\
$^{3}$ Janusz Gil Institute of Astronomy, University of Zielona G\'ora, Lubuska 2, 65-265, Zielona G\'ora, Poland \\
$^{4}$ European Southern Observatory, Karl Schwarzschild-Str. 2, D-85748, Garching, Germany \\
$^{5}$ Institute of Space Sciences (IEEC--CSIC), Carrer de Can Magrans s/n, E-08193 Barcelona, Spain \\
$^{6}$ Anton Pannekoek Institute for Astronomy, University of Amsterdam, Postbus 94249, NL-1090-GE Amsterdam, The Netherlands 
}

\begin{document}

\date{Accepted 2017 September 29. Received 2017 September 29; in original form 2017 August 4}

\pagerange{\pageref{firstpage}--\pageref{lastpage}} \pubyear{2017}

\maketitle

\label{firstpage}

\begin{abstract}
We present deep observations of the field of the magnetar \cxo\ in the star cluster Westerlund 1, obtained in the near-infrared with the adaptive optics camera NACO@VLT.  
We detected a possible candidate counterpart at the {\em Chandra} position of the magnetar, of magnitudes $\mathrm{J} = 23.5 \pm 0.2$, $\mathrm{H} = 21.0 \pm 0.1$, and  
$\mathrm{K}_\mathrm{S} = 20.4 \pm 0.1$.  The K$_{\rm S}$-band measurements available for two epochs (2006 and 2013) do not show significant signs of variability but only 
a marginal indication that the  flux varied (at the 2 $\sigma$ level), consistent with the fact that the observations were taken when \cxo\ was in quiescence. 
At the same time, we also present colour--magnitude and colour--colour diagrams in the J, H, and K$_{\rm S}$ bands from the 2006 epoch only, the only one with observations 
in all three bands, showing that the candidate counterpart lies in the main bulk of objects describing a relatively well--defined sequence. Therefore, based on its colours 
and lack of variability, we cannot yet associate the candidate counterpart to \cxo. Future near-infrared observations of the field, following-up a source outburst, would 
be crucial to confirm the association from the detection of near-infrared variability and colour evolution.
\end{abstract}


\section{Introduction}

 The {\em Chandra} X-ray source \cxo\ was discovered in the massive star cluster Westerlund 1 (Muno et al.\ 2006) as a bright X-ray source, 
modulated at a period of 10.6 s identified as the spin period $P_{\rm s}$ of an isolated neutron star (INS), as suggested by the lack of a bright
infrared counterpart.  
The period value is typical of that of a peculiar class of INSs, the magnetars, of which about 30 are known to date, as listed in the Mc Gill 
Magnetar Catalogue (Olausen \& Kaspi 2014)\footnote{http://www.physics.mcgill.ca/$\sim$pulsar/magnetar/main.html}.  Magnetars are thought to be 
INSs with enormous magnetic fields in excess of $10^{14}$ G, hence the name, which are powered by magnetic energy instead of the star rotation 
(see Kaspi \& Beloborodov 2017, for a recent review).  The blackbody (BB) X-ray spectrum ($kT=0.61$ keV) of \cxo, the  size of the emitting 
radius ($r_{\rm BB} \sim 0.3$ km), the X-ray luminosity ($L_{\rm X} \sim 3 \times 10^{33}$ erg s$^{-1}$), in excess of the pulsar rotational 
energy loss $\dot{E}\sim 3\times 10^{31}$  erg s$^{-1}$, as inferred from the period derivative $\dot{P}_{\rm s}$ (Israel et al.\ 2007; 
Rodriguez Castillo et al.\ 2014), the inferred dipolar magnetic field of $\sim 10^{14}$ G, are also typical of magnetars.  The detection 
of a bright X-ray outburst  (Krimm et al.\ 2006) is also in line with the classical magnetar behaviour. The outburst decay has been followed 
through the years with all X-ray observing facilities (Israel et al.\ 2007; Naik et al.\ 2008; Woods et al.\ 2011;  An et al.\ 2013) and is 
consistent with trends observed in other magnetars. The association of \cxo\ with Westerlund 1 has suggested that the magnetar progenitor
 was an extremely massive star of mass $\ga 50 M_{\odot}$ (Muno et al.\ 2006; Belczynski \& Taam 2008), more massive than expected according 
to INS formation theories that predict progenitors of mass $\approx 10 M_{\odot}$. It has also been suggested that \cxo\ possibly formed in 
a binary system, now disrupted after the progenitor star went supernova, with the  star Wd1-5 being the pre-supernova companion (Clark et al.\ 2014). 

Very little is known about the optical and infrared (IR) emission properties of magnetars and the underlying emission mechanisms 
 (e.g., Mignani 2011; Mereghetti 2011) .  Indeed, magnetars are elusive targets in the optical/IR, owing to their usually large 
distances and interstellar extinction, and 
only  nine of them have been  identified in the optical and/or near-IR out of the 30 currently known  (Olausen \& Kaspi 2014) 
with a flux measurement in at least one filter. Their magnetar IR emission is probably non-thermal in origin with an higher emission 
efficiency with respect to rotation-powered pulsars, possibly related to their much higher magnetic fields (Mignani et al.\ 2007). Emission 
from a debris disc might explain the magnetar emission in the mid-IR (Wang et al.\ 2006a).
No optical/IR counterpart has been detected for \cxo\  from observations carried out soon after the magnetar discovery (Muno et al.\ 2006; 
Wang et al.\ 2006b). No other IR observations of this magnetar have been reported ever since.  

Here, we present the results of near-IR observations of the magnetar carried out in two different epochs with the Very Large Telescope (VLT).  
This manuscript is structured as follows: observations and data analysis are presented in Sectn.\, 2, whereas the results are presented in 
Sectn.\, 3 and 4, respectively.

\begin{figure*}
\includegraphics[width=14cm]{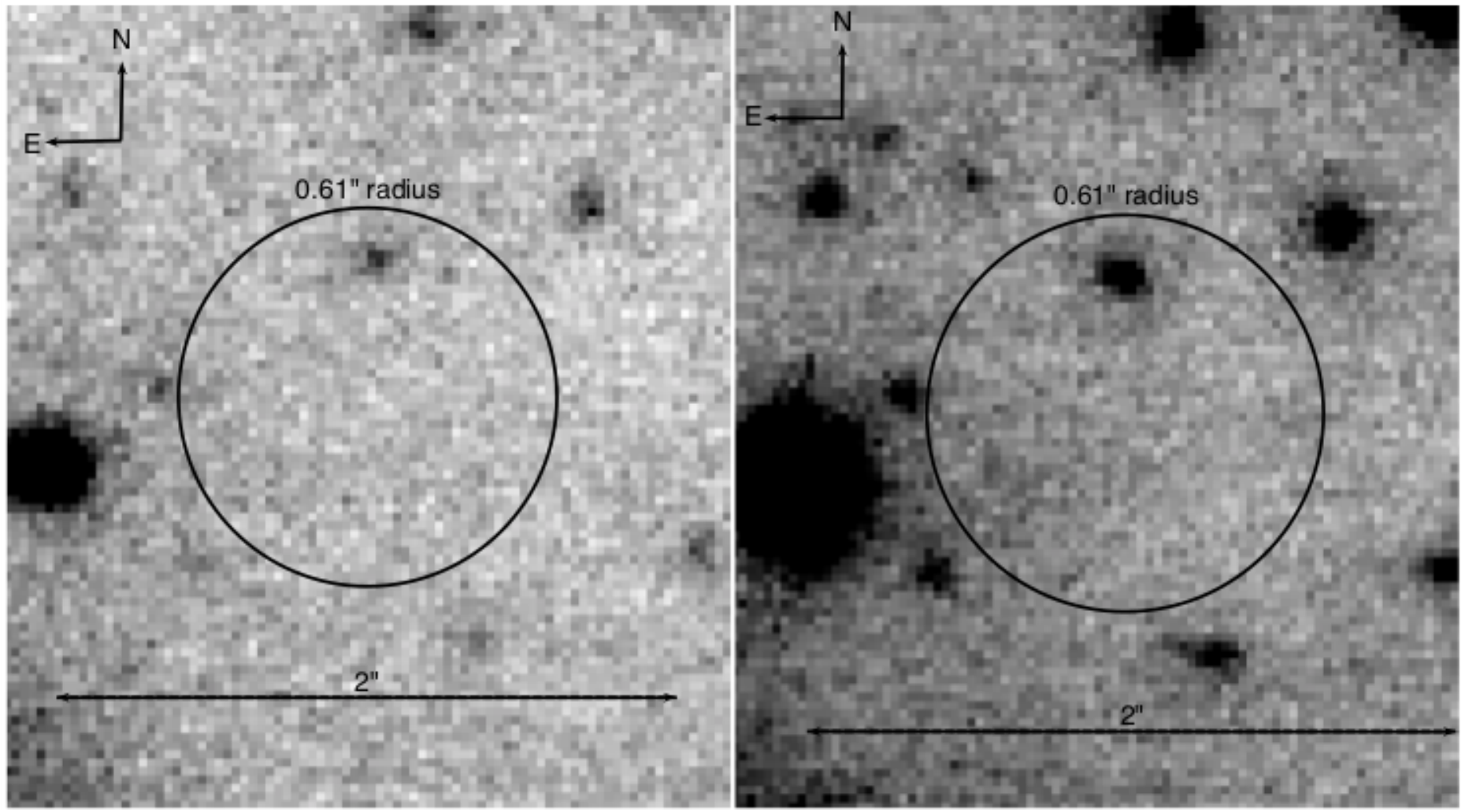}
\caption{Zoom of the \cxo\ field around its {\em Chandra} position taken in the K$\mathrm{_S}$-band at two different epochs. 
Left: June 2006, right: August 2013. The circle is drawn at the Chandra position of \cxo\ and has radius equal to the overall 90\% confidence level position uncertainty (0\farcs61).}
\label{map}
\end{figure*} 

\section{Observations, Reductions and Calibrations}

The data-set consists of two-epoch observations of the field of \cxo\ obtained with the VLT@ the ESO--Paranal Observatory (Chile), 
equipped with the  NAOS-CONICA Adaptive Optics (AO) camera (NACO; Lenzen et al.\ 2003; Rousset et al.\ 2003). In both epochs, the 
NACO S27 camera  was used, with a field--of--view of $27\arcsec\times 27\arcsec$ and a scale of 0\farcs027/pixel. We retrieved the 
first epoch data from the public ESO archive\footnote{www.eso.org.archive}, while the second
epoch data were obtained via a dedicated proposal submitted by our group (Programme ID: 091.D-0071(B), PI Mignani). 
For the first epoch --June 26 and 27, 2006,  images were taken in  the three wide bandpass J, H, K$_{\mathrm{S}}$ filters, 
while in the second epoch -- August 16, 2013, only the K$_{\mathrm{S}}$ filter was used. Images were taken using the Double-Correlated 
read-out mode with several single dithered frames (DIT) repeated n-times (NDIT) and stacked together along each node of the dithering
pattern. The choice of DIT and NDIT depends on the used filter and the epoch. In particular, DIT is usually shorter in the long wavelength 
filters  because of the higher background level. 
For the first epoch data-set a total of 2520, 2070 and 1870 s of integration time was obtained
for the J, H, and K$_{\mathrm{S}}$ filters, respectively, while for the second epoch one a total of 7425 s of integration time 
in the K$_{\mathrm{S}}$ was secured. For each epoch, and each filter, single images were pre-reduced (flat fielding, sky subtraction, 
exposure map correction) and co-added through the NACO pipeline obtaining a master image. 
The full width half maximum (FWHM) of a point source  computed on each master image is  0\farcs123, 0\farcs106 and 0\farcs110 for 
the J, H and K$_{\mathrm{S}}$ filters of the first epoch, and 0\farcs116 for the K$_{\mathrm{S}}$ of the second epoch. 

Object photometry was performed by means of the widely--used package {\tt DAOPHOT-II} (Stetson 1994) that we used to obtain 
multi-band catalogues of all objects found in the NACO images down to a detection limit of $\sim 5 \mathrm{\sigma}$ over the 
background. We calibrated our object catalogues via comparison with the 2MASS Point Source Catalogue (Skrutskie et al.\ 2006), 
that was also used to compute the astrometric solution for the NACO images. This last process presented some difficulties because 
it may happen that a single 2MASS star is actually  resolved in a multiple object on the almost diffraction-limited NACO images, 
so that the astrometric calibration has to be performed carefully to prevent systematics. 
Moreover, in a field-of-view as small as that of NACO, only very few 2MASS stars are present, and generally affected by strong 
saturation and/or non-linearity problems. In order to by-pass this problem, we computed the astrometric calibration via a two-step 
process using an archival K-band image of the field taken with the  Son Of Isaac (SOFI) infrared Camera at the ESO New Technology 
Telescope (NTT).  As a first step we calibrated the astrometry of the SOFI image, of area $\approx 4\arcmin \times 4\arcmin$, by 
using the 2MASS and UCAC4  (Zacharias et al.\ 2013) catalogs. Then, we used a catalog of secondary stars detected on the SOFI image, 
fainter than the 2MASS stars and with many more matches on the NACO images, to derive their final astrometric solution. In this way, 
we obtained an astrometric solution with an overall uncertainty of about 87 mas, coming from the error propagation on the differences 
in position in the sense (found-reference) to which the high-end of the overall precision figure of the 2MASS survey 
(80 mas\footnote{www.ipac.caltech.edu/2mass/releases/allsky/doc/sec2\_2.html}) has been added in quadrature.
Analogously, photometric calibration via 2MASS shows the same potential problem, and in case a 2MASS star is resolved in a multiple 
object, a correct calibration is obtained by summing up the fluxes of all the stars that appear as a single blended object in the 2MASS catalogue. 
In order to circumvent this problem, and considering also that the bright 2MASS stars are in general non-linear in the 
NACO images, we adopted the two-step calibration via SOFI images also for the photometric calibration. At the end of the process, the obtained 
zero points have an uncertainty of $\sim 0.1$ magnitudes in all three filters and two epochs. 

Since we are interested in detecting a possible variability of the candidate counterpart(s) to \cxo, which is traditionally 
considered a marker of the identification of a magnetar counterpart, e.g., Tam et al.\ 2004,  particular attention has 
been devoted to relative calibration of one,data-set against the other in the only filter for which we had two-epochs data, the  K$_{\mathrm{S}}$ one. 
In this way, glitches and systematics possibly affecting the absolute photometric calibration process are strongly reduced. 
We remind, however, that the lack of infrared variability, correlated or not with variations in the X-ray flux (see, e.g., Durant \& van Kerkwijk 2006), 
does not necessarily argue against a potential counterpart identification. 
For instance, in the case of the infrared counterpart to the magnetar XTE\, J1810$-$197 variations of the infrared flux 
do not follow the same trend observed in the X-ray (Testa et al.\ 2008).

  \begin{figure*}
 \centering
\parbox{8cm}{
\includegraphics[width=8cm]{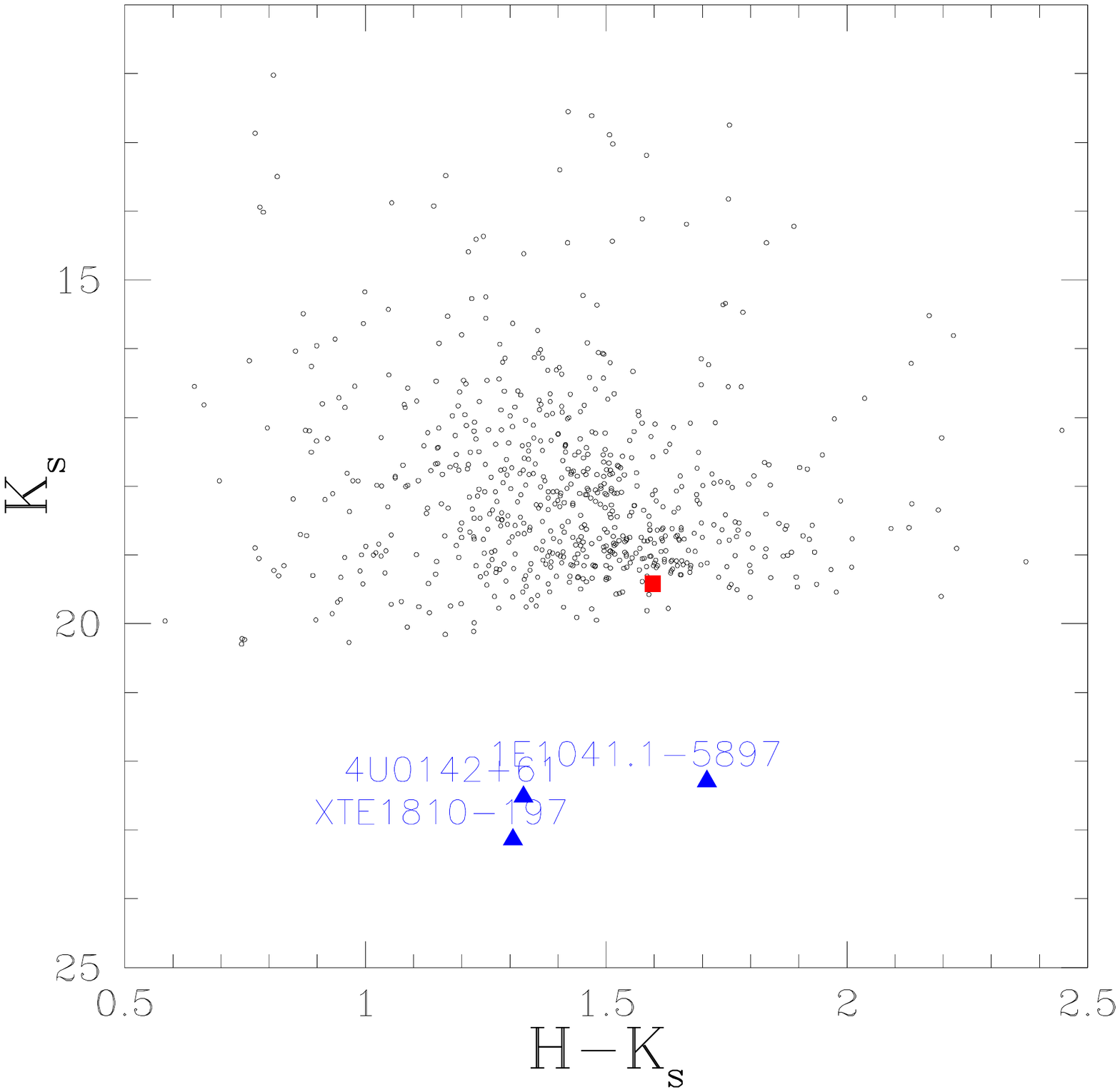}}
\qquad
\parbox{8cm}{
\includegraphics[width=8cm]{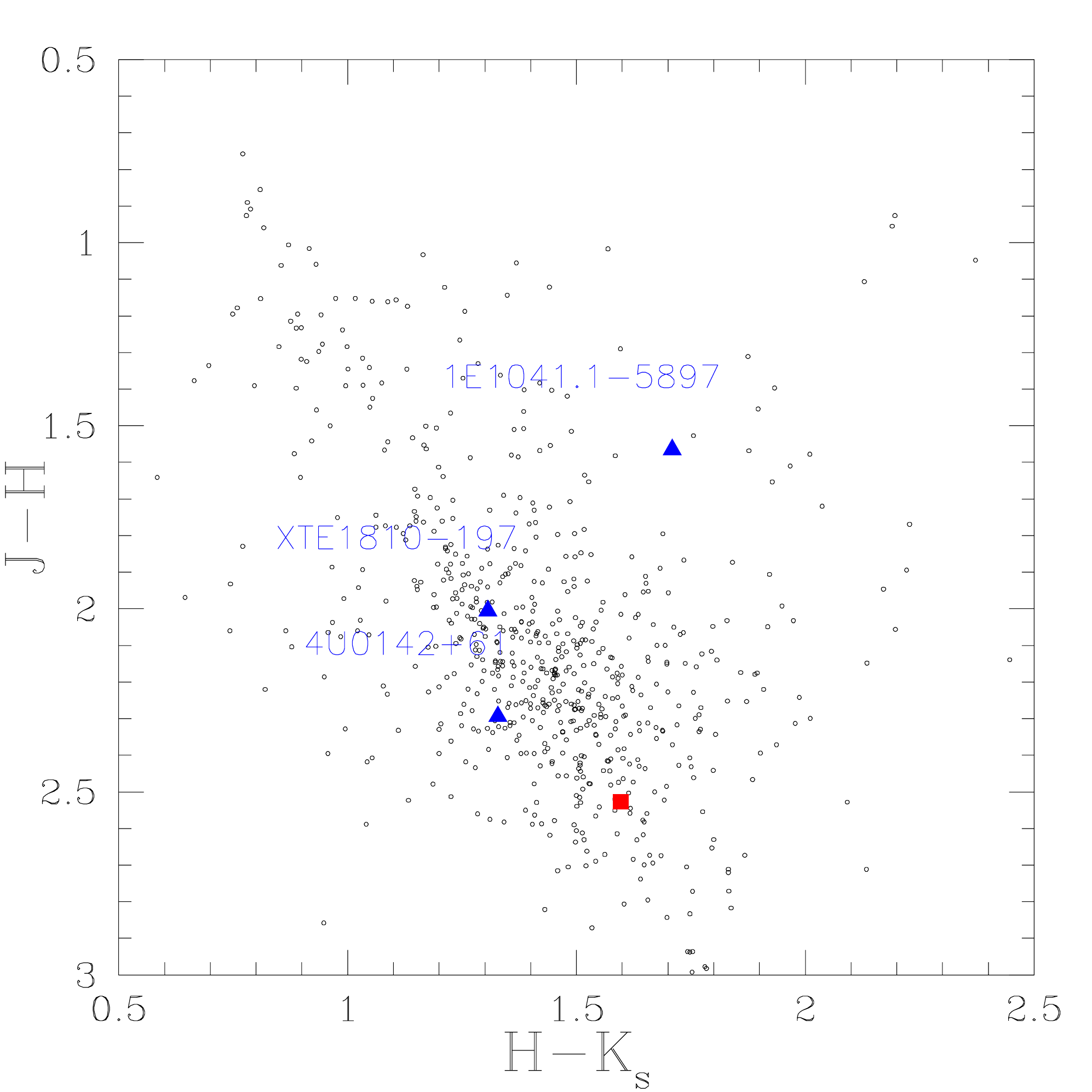}}
\caption{Colour-magnitude (left) and colour-colour (right) diagrams for all sources in the \cxo\ NACO field--of--view. The candidate counterpart is marked with a red square dot. 
The blue triangles correspond to magnetars with magnitudes in all the three near-IR filters, plotted as a reference. 
Their observed fluxes have been scaled according to the nominal reddening and distance values. All values are from the Mc Gill Magnetar Catalogue and refs. therein.}
\label{cmd}
\end{figure*} 

\section{Near-infrared candidate counterpart}

Figure \ref{map} shows a $\sim$3$\arcsec \times$3$\arcsec$ map of an area centred on the nominal coordinates of \cxo. The different depth due to the much longer exposure 
time of the second-epoch image can be appreciated by eye. 
In both maps, circles are drawn at the nominal \cxo\ position, with radius corresponding to the 90\% positional uncertainty of the X-ray source. 
This has been calculated by adding in quadrature the uncertainty on the primary astrometric calibrators (2MASS, 80mas - we adopt the higher end of the 
range given for the catalog), the internal r.m.s. of the astrometric fit, which is considerably smaller (0.4 pixels = 0\farcs011 or 11 mas), and the 
uncertainty on the \cxo\ position. We computed the coordinates of \cxo\ from the available {\em Chandra} observations of the field. In particular, 
we used three ACIS-S observations (Obs Id.: 5411, 6283, 14360) taken in imaging mode where the pulsar has always been observed on axis and located on 
the back-illuminated chip S3, which is optimised for source positioning. 
We derived average coordinates of  $\alpha =16^{\rm h}  47^{\rm m} 10\fs21$; $\delta  = -45^\circ 52\arcmin 17\farcs06$, 
where the statistical error on the position is negligible with respect to the systematic uncertainty on the absolute {\em Chandra} astrometry, 
which is 0\farcs6 at the 90\% confidence level\footnote{{\texttt http://cxc.harvard.edu/ciao/ahelp/coords.html}}. 
An attempt was made to improve the accuracy of the {\em Chandra} astrometry and to reduce the uncertainty on the position of our target by using 
other sources detected in the ACIS-S chip 3 field of view. We found several 
X-ray sources in the {\em Chandra} images of \cxo.  However, the aim point of these images is the core of the Westerlund 1 cluster, where the high density 
of objects makes it extremely difficult to associate an X-ray source with a high-confidence candidate counterpart, even at the {\em Chandra} spatial 
resolution. Indeed, all our attempts resulted in a large scatter around the best-fit match between the X-ray coordinates and the optical/infrared ones, 
comparable to (or larger than) the nominal 90\% {\em Chandra} absolute astrometry accuracy.
Therefore, a conservative approach was adopted and the above value was used.
We note that the overall uncertainty of our astrometry is much larger than any reasonable figure of the positional uncertainty due to an unknown proper 
motion of the source.  For instance, assuming an average neutron star velocity of 400 km s$^{-1}$ (Hobbs et al.\ 2005) and a distance of 5 kpc would 
give a figure of 0\farcs016 yr$^{-1}$.

As can be seen from the map(s) in Figure \ref{map}, one object clearly falls within the error circle
There are other faint fluctuations visible within the error circle, but those have marginal significance, around or below 1 $\sigma$ and
are undetected by the source detection algorithm of {\tt DAOPHOT}, even when pushed to very faint detection limits, 
and we regard them as background fluctuations, although they might deserve further attention should any deeper observation ever be carried out in the future.
 As far as our results show, we assume the object in the error circle as the candidate IR counterpart of \cxo.  Its magnitudes are $\mathrm{K}_\mathrm{S} = 20.4 \pm 0.1$ and $\mathrm{K}_\mathrm{S} = 20.0 \pm 0.1$ in the 2013 and 2006 images, respectively.
For the J and H filters, available as we said for the first epoch only, we find: $\mathrm{J} = 23.5 \pm 0.2$ and $\mathrm{H} = 21.0 \pm 0.1$ 
No other object has been detected down to 3$\sigma$ magnitude limits of 23.6, 22.3, 21.1 in J, H, K$_{\rm S}$, respectively (first epoch), whereas the K$_{\rm S}$-band limit for the second epoch is deeper, namely K$_{\rm S}$ = 21.9.

Given the large number of stars in the crowded Westerlund 1 field the chance coincidence probability to find an object within a radius of 0\farcs61 is quite high.   The probability $P=1-\exp(-\pi\rho r^2)\sim90\%$, where $r=0\farcs61$ 
and $\rho\sim 2$ arcsec$^{-2}$ is  the number density of objects in the NACO field--of--view measured in the  K$_{\rm S}$-band image.  However, we cannot rule out a priori that this object is, indeed, associated with \cxo.

Our candidate might be one of the "two very faint objects" found near the X-ray source position in observations obtained  on May 17, 2006 with PANIC (Persson's Auxilliary Nasmyth Infrared Camera) at the 6.5 m  
Magellan-Baade telescope (Las Campanas) by Wang et al.\ (2006b), when the source was in quiescence.  These "two very faint objects" were not detected in PANIC follow-up observations of the \cxo\ X-ray outburst (Krimm et al.\ 2006) 
taken on September 29, 2006  (Wang et al.\ 2006b). This suggested that neither of them could have been the magnetar counterpart, for which a brightening of $\sim$ 4.6 magnitudes would have been expected if the infrared flux of the 
magnetar scales exactly as the X-ray flux (e.g. Tam et al.\ 2004). In this case, the ratio between the observed X-ray flux in outburst and in quiescence (Muno et al.\ 2006) would translate into the expected variation of 4.6 magnitudes 
in the infrared flux (Wang et al.\ 2006b). However,  Wang et al.\ (2006b)  did provide neither a finding chart nor a reference magnitude, so that we cannot tell if our candidate counterpart is one of 
their "two very faint objects" and any comparison with their results would be, at this point, just speculative.
Our limits are slightly fainter than those obtained by Wang et al.\ (2006b, K$_\mathrm{S} \sim$  21) for the 2006 data and one magnitude fainter for the 2013 ones.
Figure \ref{cmd} shows the colour--magnitude (CMD) and colour--colour (CCD) diagrams of the catalog of objects in common between the two epochs. The position of the \cxo\ candidate counterpart is highlighted with a larger, red square dot.The candidate has  magnitudes and colours of the bulk of stars redder than the main sequence that might belong to the field population (see Kudryavtseva et al.,\ 2012 for an analysis of the cluster population), despite what would be expected if it was the actual counterpart of \cxo.  For comparison, Figure \ref{cmd}  also shows the observed colours for a sample of magnetars with detections in the J, H, and K$_{\rm S}$ bands, where the flux values have been taken from the Mc Gill Magnetar catalogue. As seen, at variance with the \cxo\ candidate counterpart, the colours of the magnetars counterparts are quite bluer from those of a field stellar population, 
not to mention the fact that the magnitudes of known magnetars, scaled to the distance and reddening of Westerlund 1 are fainter than our candidate. 
However, very little is still  known on the magnetars's near-IR spectra and on their evolution with the source state, which means that assuming magnetar colours derived from flux measurements that might have been obtained at different epochs as a template has to be done with due caution. Indeed, our first-epoch NACO observations were taken within one day from each other and when the source was in quiescence (see below). Moreover, magnetars' luminosities in the near-IR can differ by three orders of magnitude (e.g., Mignani et al.\ 2007), so that none of them can be assumed as a 'standard candle'.

\begin{figure}
\includegraphics[width=8cm]{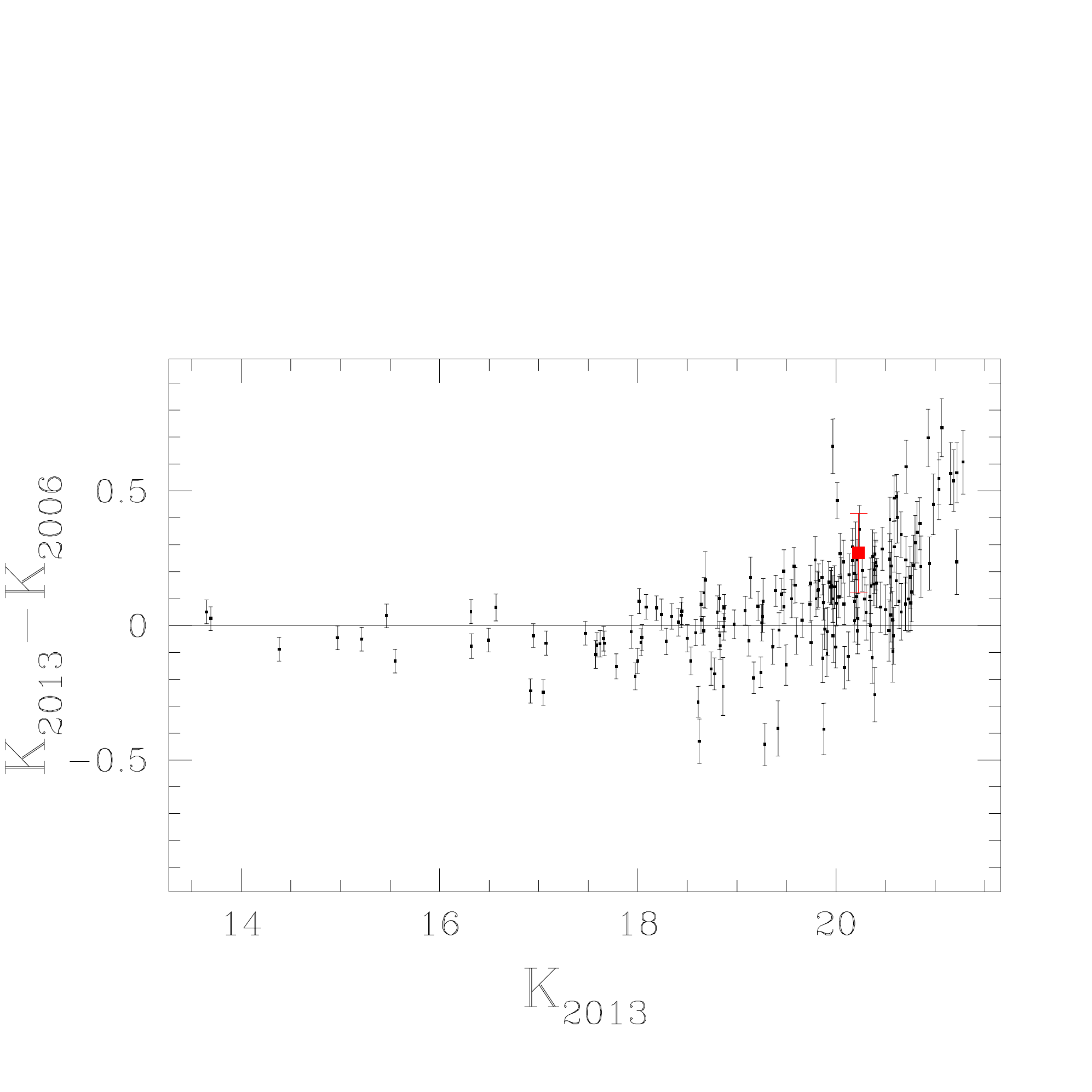}
\caption{Difference between the K$\mathrm{_S}$-band magnitudes of the \cxo\ candidate counterpart  as measured in the first (June 2006) and second (August 2013) epochs of the NACO observations. The red point corresponds to  the candidate counterpart, whereas the black dots corresponds to field stars selected in the object vicinity within 300 pixels and with good photometry.}
\label{variability}
\end{figure}

It is interesting to note that both NACO  observations were carried out while the source was in quiescence, since the first epoch (June 2006) falls before the outburst detected in September 2006 (Krimm et al.\ 2006), and the second epoch one (August 2013) falls after another less intense burst that has been registered in November 2011. No other outbursts have been registered ever since (see Rodrigurez Castillo et al.\ 2014, and refs. therein for details), till the very recent one occurred on May 16, 2017 (D'Ai et al.\ 2017).
Figure \ref{variability} shows a comparison plot of the 2013 vs 2006 K$_\mathrm{_S}$-band magnitudes for a number of objects in a restricted area around the position of \cxo , aimed at studying the variability of its candidate counterpart. The plot shows signs of variation of the candidate counterpart, but they are well within the uncertainties at about a $2 \sigma$ level, consistent with the hypothesis that the source did not vary its IR flux from the first to the second of the two epochs.  
This is not surprising if one takes into account the quiescence of the magnetar at both epochs. 
If the object that we detected in the {\em Chandra} error circle of \cxo\ was its actual counterpart it would have a K$_{\rm S}$-band luminosity of $\sim 1.12 \times 10^{31}$ erg s$^{-1}$, at the Westerlund 1 distance of 5 kpc and assuming the K$_{\rm S}$-band interstellar 
extinction from the Mc Gill Magnetars' catalogue ($A_{\rm K}=1.4$). This luminosity would be about 37\% of the pulsar rotational energy loss, a quite high fraction but still comparable to those of other magnetars and in line with the \cxo\  magnetic field of $10^{14}$ G (Mignani et al.\ 2007). The observed K$_{\rm S}$-band--to--X-ray flux ratio would be $\sim 6.1 \times 10^{-3}$, where the X-ray flux in quiescence has been measured in the 1--10 keV energy band, still in the range typical of magnetars (Rea et al.\ 2010). Therefore, at least according to its expected near-IR luminosity and near-IR/X-ray emission, the source that we detected in the \cxo\ error circle would still be a plausible candidate counterpart.

\section{Summary and Conclusions}
In this study we present an attempt to identify the near-IR counterpart to the magnetar \cxo\ by means of AO observations obtained in two epochs, the first one before the outburst of 2006 (Krimm et al.\ 2006) and the second one well after the second outburst in 2011. We identify a main candidate within the 90\% confidence level error circle around the nominal position of \cxo.
About the candidate, it shows no obvious variability between the two epochs, but this is an expected result since the X-ray source was in a quiescent state at both epochs. Moreover, the candidate has the characteristics of a main sequence star showing no particular colours, which would be expected for a magnetar. Monitoring the evolution of the source brightness in the near-IR, following outburst activities in the X-ray, would be crucial to determine whether this source is the actual counterpart to \cxo. 
Nevertheless, new deeper observations with AO-equipped large telescopes, e.g. the forthcoming ELTs, are needed to confirm if the candidate is really the magnetar counterpart and, in case of success, determine its nature.

\section*{Acknowldegements}
We thank the anonymous referee for his/her considerate and constructive review of our manuscript. RPM acknowledges financial support from an "Occhialini Fellowship".

\end{document}